\documentclass[twocolumn,noshowpacs,preprintnumbers,amsmath,amssymb,showkeys]{revtex4}

\usepackage{graphicx}
\usepackage{dcolumn}
\usepackage{bm}

\begin{document}

\preprint{Preprint 1/2006}

\title{Least Squares Importance Sampling for Monte Carlo Security Pricing}

\author{Luca Capriotti} \email{luca.capriotti@credit-suisse.com.}

\affiliation{%
Global Modelling and Analytics Group, Investment Banking Division, Credit Suisse Group\\
One Cabot Square, London, E14 4QJ, United Kingdom
}%


\date{\today}

\begin{abstract}
We describe a simple Importance Sampling strategy for Monte Carlo
simulations based on a least squares optimization procedure. With
several numerical examples, we show that such Least Squares
Importance Sampling (LSIS) provides efficiency gains comparable to
the state of the art techniques, when the latter are known to
perform well. However, in contrast to traditional approaches, LSIS
is not limited to the determination of the optimal  mean of a
Gaussian sampling distribution. As a result, it outperforms other
methods when the ability to adjust higher moments of the sampling
distribution, or to deal with non-Gaussian or multi-modal
densities, is critical to achieve variance reductions.
\end{abstract}

\maketitle

\section{Introduction}

The impressive development of the securities markets has generated
in the last few years a steady demand for more and more structured
financial products. At the same time, the level of sophistication
and complexity of the pricing models employed by investment firms
has dramatically increased, in a continuous
search for a possible edge against competitors.  As a result, an
increasing fraction of the models employed in practice is
too complex to be treated by analytic or deterministic numerical
methods (trees or partial differential equations), and Monte Carlo
simulation becomes more often than ever the only computationally
feasible means of pricing and hedging.

Although generally easy to implement, Monte Carlo simulations are
infamous for being slow. In fact, being stochastic in nature,
their outcome is always affected by a statistical error, that can
be generally reduced  to the desired level of accuracy by
iterating the calculation for long enough time.  This comes with a high computational cost
as such statistical uncertainties, all things being equal, are
inversely proportional to the {\rm square root} of the number of
statistically independent samples. Hence, in order to reduce the
error by a factor of 10 one has to spend 100 times as much
computer time. For this reason, to be used on a trading floor,
Monte Carlo simulations often require to be run on large parallel
computers with a high financial cost in terms of hardware,
infrastructure, and software development.

Motivated by this very practical necessity, several approaches to
speed up Monte Carlo calculations, such as Antithetic Variables,
Control Variates, and Importance Sampling, have been proposed over
the last few years \cite{GlassMCbook}. These techniques aim to
reduce the variance per Monte Carlo observation so that a given
level of accuracy can be obtained with a smaller number of
iterations. In general, this can be done by exploiting some
information known {\em a priori} on the structure of the problem at
hand, like a symmetry property of the Brownian paths (Antithetic
Variables), the value of a closely related security (Control
Variates), or the form of the statistical distribution of the
random samples (Importance Sampling). Antithetic Variables and
Control Variates are the most commonly used variance reduction
techniques, mainly because of the simplicity of their
implementation, and the fact that they can be accommodated  in an
existing Monte Carlo calculator with a small effort. However,
their effectiveness varies largely across applications, and is
sometimes rather limited \cite{GlassMCbook}.

On the other hand, Importance Sampling techniques, although
potentially more powerful, have not been employed much in
professional contexts until recently.  This is mainly
because such techniques generally involve a bigger implementation
effort, and they are also less straightforward to include in a
general Monte Carlo framework. Moreover, when used improperly,
Importance Sampling can increase the variance of the Monte Carlo estimators, 
thus making its
integration in an automated environment more problematic.
Nonetheless, the potential efficiency gains at stake are so large
that the interest in finding efficient Importance Sampling schemes
is still very high.

The idea behind Importance Sampling is to reduce the statistical
uncertainty of a Monte Carlo calculation by focusing on the most
important sectors of the space from which the random samples are
drawn.   Such regions critically depend on both the random
process simulated, and  the structure of the security priced.
For instance, for a deep out-of-the money Call option \cite{Hull},
the payoff sampled is zero for most of the iterations of a Monte
Carlo simulation. Hence, simulating more samples with positive
payoff reduces the variance. This can be done by changing the
probability distribution from which the samples are drawn, and
reweighing the payout function by the appropriate likelihood-ratio
(Radon-Nikodyn derivative) in order to produce an unbiased result
of the original problem \cite{GlassMCbook}.

Most of the work in Importance Sampling methods for security pricing has been done in a
Gaussian setting
\cite{reider,BoyleBroadieGlass97,VasDuf98,GlassImportSampl99,
GlassImportSampl99hjm,SUFU00,SUFU02,Arouna03} such the one arising
from the simulation of a diffusion process. In this framework,
Importance Sampling is achieved by modifying the drift term of the
simulated process in order to drive the Brownian paths towards the
regions that are the most important for the evaluation of the
security. For instance, for the  Call option above, this can be
obtained by increasing the drift term up to a certain
optimal level \cite{reider,BoyleBroadieGlass97}. The different
approaches proposed in the literature, essentially differ in the
way in which such change of drift is found, and can be roughly
divided into two families depending on the strategy adopted. The
first strategy, common to the so-called adaptive Monte Carlo
methods \cite{VasDuf98,SUFU00,SUFU02,Arouna03}, aims to determine
the optimal drift  through stochastic optimization techniques that
typically involve an iterative algorithm. On the other hand, the
second strategy, proposed in a remarkable paper by Glasserman,
Heidelberger, and Shahabuddin (GHS) \cite{GlassImportSampl99},
relies on a deterministic optimization procedure that can be
applied for a specific class of payouts. This approach, although
approximate, turns out to be very effective for several pricing
problems, including the simulation of a single factor
Heath-Jarrow-Morton model \cite{GlassImportSampl99hjm}, and
portfolio credit scenarios \cite{GlassLi05}.

In this paper, we propose the Least Squares Importance Sampling
(LSIS), as an effective and flexible variance reduction method for
Monte Carlo security pricing. This approach, originally proposed in Physics
for the optimization of quantum mechanical wave functions of
correlated electrons \cite{umrigar98}, is applied here in a
financial setting. In LSIS the determination of the optimal drift
-- or more in general of the most important regions of the sample
space -- is formulated in terms of a least squares minimization. 
This technique can be easily implemented and included
in an existing Monte Carlo code, and simply relies on a
standard least square algorithm for which several optimized
libraries are available. We will show that LSIS provides
efficiency gains analogous to the ones of previously proposed
methods when the latter are known to perform well. However, LSIS
does not share some of their limitations, and its range of
applications includes multi-modal problems and non-Gaussian
sampling.

In the following Section, we begin by reviewing the main ideas
behind Importance Sampling in a generic Monte Carlo framework.
Then in Section III we specialize the discussion to the Gaussian
setting, and we briefly review the recent adaptive strategies, and
the GHS approach of Ref.\cite{GlassImportSampl99}.
The rationale of LSIS is introduced in Section IV
together with the essential implementation details, and in Section
V we present the results of our numerical experiments. Here we
perform a systematic study of the variance reductions obtained by
means of LSIS for a variety of test cases, including a comparison
with recent Importance Sampling techniques. Finally, we draw our
conclusions in Section VI.

\begin{figure}
\vspace{-14mm}
\includegraphics[width=0.43\textwidth]{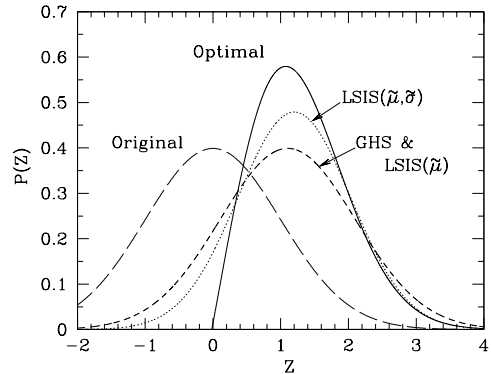}
\vspace{-10mm} \caption{\label{figcall} Sampling probability
density functions for a European Call option with $T = 1$, $r =
0.05$, $\sigma = 0.3$, $X_0 = K = 50$ as obtained with LSIS
[optimizing just the drift, LSIS$(\tilde\mu)$, and both the drift and
the volatility, LSIS$(\tilde\mu,\tilde\sigma)$], and the saddle point approximation
(GHS). On this scale the results for LSIS$(\tilde\mu)$ and GHS are
indistinguishable. The original (\ref{gaussmulti}) and the optimal
(\ref{zerovar}) sampling densities are also shown for comparison.
}
\end{figure}

\section{Importance Sampling}
\label{is}

 Let us consider the general problem of estimating the
expectation value of a scalar function, $G(Z)$, depending on a
$d$-dimensional real random vector $Z = (Z_1,\ldots,Z_d)$ with
joint probability distribution $P(Z)$,
\begin{equation}\label{expectproblem}
V = {E}_P\left[ G(Z) \right] = \int_D \hspace{-1mm} dZ \,\, G(Z)\,
P(Z)~,
\end{equation}
where $D$ is the domain of possible values of the state variables
$Z$ \footnote{In the present discussion we will treat the
$Z_i$'s as continuous variables, however all the results also
apply if some or all of them can assume only a discrete set of
values. For any of such variable, the symbol $\int dZ_i$ is to be
interpreted as a sum over the possible outcomes.}.
In a financial derivatives context, $G(Z)$ would typically
represent the discounted payout of a certain security, and $P(Z)$
would be the risk neutral-probability measure of an arbitrage free
market \cite{HarrKreps,Musiela,BaxterRennie}. For instance, for
the familiar Call option in the Black-Scholes framework
\cite{Hull} one has $d=1$, $P(Z) = (2\pi)^{-1/2}
\exp{(-Z^2/2)}$ and
\begin{equation}
G(Z) =
e^{-rT}\left(X_0\exp{\left[\left(r-\frac{\sigma^2}{2}\right) T +
\sigma Z\right]}-K\right)^+
\end{equation}
 where $r$ is the risk-free interest rate, $\sigma$ is the
volatility, $X_0$ and $K$ are respectively the spot and strike
price, and $T$ the maturity of the option.

Whenever the dimension $d$ of the state variable $Z$ is large (say
$d \gtrsim 4$) standard numerical quadrature approaches become
highly inefficient, and Monte Carlo methods are the only feasible
route for estimating expectation values of the form
(\ref{expectproblem}). To do so, one interprets
Eq.~(\ref{expectproblem}) as a weighted average of the payout
function $G(Z)$ over the possible configurations $Z$ with weights
given by the probability distribution $P(Z)$. This immediately
leads to the simplest (and crudest) Monte Carlo estimator which is
obtained by averaging the payout function over a sample of $N_p$
{\em independent} values of the random variable $Z$ generated
according to the probability distribution $P(Z)$,
\begin{equation}\label{crude}
V \simeq \bar V = \frac{1}{N_p} \sum_{i=1}^{N_p} G(Z_i) ~~~~~~ Z_i
\sim P(Z)~.
\end{equation}
In particular, the central limit theorem \cite{CLT} ensures that,
for big enough samples, the values of the estimator $\bar V$ are
normally distributed around the true value, and converge for $N_p
\to \infty$ towards $V$ namely
\begin{equation}\label{sqrt}
V \simeq \frac{1}{N_p} \sum_{i=1}^{N_p} G(Z_i) \pm
\frac{\Sigma}{\sqrt{N_p}}~,
\end{equation}
where $\Sigma^2 =
E_P\left[G(x)^2\right]-E_P\left[G(x)\right]^2$ is
the variance of the estimator and can be similarly
approximated by
\begin{equation}
\Sigma^2 \simeq  \frac{1}{N_p} \sum_{i=1}^{N_p} \left(G(Z_i)
- \bar V \right)^2~.
\end{equation}
Although Eq.~(\ref{sqrt}) ensures the convergence of the Monte
Carlo estimator to the expectation value (\ref{expectproblem}),
its practical utility depends on the magnitude of the variance,
$\Sigma^2$.
Indeed, the square root convergence in (\ref{sqrt}),
implies that the number of replications $N_p$ that are
(asymptotically) necessary to achieve a given level of accuracy is
proportional to the variance of the estimator
\footnote{In particular, the Monte Carlo integration
becomes unfeasible if the variance of the estimator diverges, 
giving rise to the so-called {\em
sign-problem} instability. Although this problem is the crux of
Monte Carlo simulations in several branches of the Physical
Sciences, see, e.g., S. Sorella and L. Capriotti, Physical Review
B {\bf 61}, 2599 (2000), this issue does not usually affect
financial contexts.}.
Roughly speaking,
such quantity is relatively small whenever the function $G(Z)$ is
approximately constant over the region of values of $Z$ that is
represented the most among the random samples, i.e., the region
that contains most of the probability mass of $P(Z)$. This is
generally not the case for most of the pricing problems
encountered in practice, and the calculation of accurate estimates
of the expectation value (\ref{expectproblem}) may require large
sample sizes $N_p$, thus becoming computationally demanding.

\begin{figure}
\vspace{-25mm}
\includegraphics[width=0.43\textwidth]{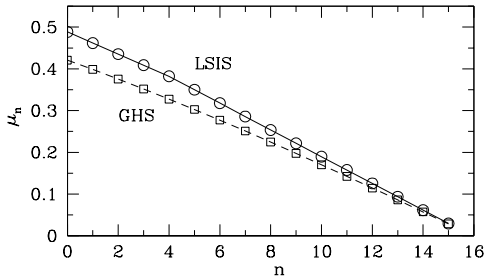}
\vspace{-10mm} \caption{\label{figmu} Optimal drift vector  as
obtained with the LSIS and the GHS procedures for an arithmetic
Asian Call option (\ref{asian}), $\Phi(\bar X) = (\bar X -K^{+})$,
on a lognormal asset (\ref{lognormal}) for $M=16$ $\sigma =0.3$,
$X_0 = K=50$, $r = 0.05$,  and $T=1.0$.}
\end{figure}

However, the choice of extracting the random variable $Z$
according to the probability distribution $P(Z)$, although
natural, is by no means the only possible one. Indeed, the Monte
Carlo integration can be performed by sampling an arbitrary
probability distribution $\tilde P(Z)$ provided that the integral
is suitably reweighed. In fact, using the identity
\begin{equation}
\int_D \hspace{-1mm} dZ \,\, G(Z)\, P(Z) = \int_D \hspace{-1mm} dZ
\,\, \frac {G(Z) P(Z)}{\tilde P(Z)}\, \tilde P(Z)~,
\end{equation}
an alternative estimator of the expectation value
(\ref{expectproblem}) is readily found as
\begin{equation}\label{isamp}
V \simeq \tilde V = \frac{1}{N_p} \sum_{i=1}^{N_p} W(Z_i) \,
G(Z_i) ~~~~~~ Z_i \sim \tilde P(Z)~,
\end{equation}
with the weight function given by $W(Z) = P(Z)/\tilde P(Z)$.
The variance of the new Monte Carlo estimator reads
\begin{equation}\label{varnew}
\tilde \Sigma^2 = \int_D \hspace{-1mm} dZ \,\, \left(W(Z)\,G(Z) -
V\right)^2\,\tilde P(Z)
\end{equation}
and critically depends on the choice of the sampling probability
distribution $\tilde P(Z)$. For non-negative functions $G(Z)$, the
optimal choice of $\tilde P(Z)$ is the one for which $\tilde
\Sigma$ vanishes, namely:
\begin{equation}\label{zerovar}
 P_{\rm opt}(Z) = \frac{1}{V} \, G(Z) P(Z)~.
\end{equation}
In fact, the Monte Carlo estimator corresponding to such {\em
optimal sampling distribution} reads
\begin{equation}
\tilde V \simeq \frac{1}{N_p} \sum_{i=1}^{N_p} W(Z_i) G(Z_i) =
\frac{1}{N_p} \sum_{i=1}^{N_p} V  ~,
\end{equation}
leading to a constant value $V$ on each Monte Carlo replication,
and resulting therefore in zero variance \footnote{It is possible to show
\cite{numrec} that, when $G(Z)$ does not have a definite sign, the
optimal sampling density has the similar form $P_{opt} = |G(Z)|
P(Z)/V$, although in this case the optimal variance is not zero.}.
Unfortunately, such a choice is not really viable as the
normalization constant, $V$, is the expectation value
(\ref{expectproblem}) we want to calculate in the first place.

\begin{table*}
\caption{\label{tablecall} Comparison between LSIS, the adaptive
Robbins-Monro (RM) algorithm (as quoted in Ref.~\cite{Arouna03}),
and the saddle point approach of Ref.~\cite{GlassImportSampl99}
(GSH): price of a European Call option on a lognormal asset
(\ref{lognormal}) for different values of the volatility $\sigma$,
and of the strike price $K$. The variance reduction (VR) is defined
in Eq.~(\ref{varred}). The parameters used are $r = 0.05$,
$X_0=50$, $T=1.0$, and the number of simulated paths is 1,000,000
for Crude MC, LSIS and GHS, 50,000 for RM. Results for LSIS
obtained by optimizing  the drift only [LSIS$(\tilde\mu)$], and
both the drift and the volatility [LSIS$(\tilde\mu,\tilde\sigma)$]
are reported. The uncertainties on the least significative digits
of the option prices, and variance reductions are reported in
parentheses. Note that in Ref.~\cite{Arouna03} no error estimate
was quoted for the RM results. However, from the ratio of the
simulated paths, it is sensible to estimate the errors on the RM
variance reductions as about $\sqrt{20} \simeq 4 $ times larger
than those quoted for LSIS$(\tilde\mu)$. }
\begin{ruledtabular}
\begin{tabular}{ccccccccccc}
        &      & Crude MC& \multicolumn{2}{c}{LSIS($\tilde\mu$)}   & \multicolumn{2}{c}{LSIS$(\tilde\mu,\tilde\sigma)$}&\multicolumn{2}{c}{RM}   &\multicolumn{2}{c}{GHS}  \\
$\sigma$ & $K$ & Price & Price    & VR      & Price        & VR        & Price & VR      & Price         & VR     \\ \hline
0.1&30& 21.4633(50) & 21.46294(49)&  104(1) & 21.46296(12) & 1700(100) & 21.47 & 112(4)  & 21.46294(50)  &100(1)  \\
&50   &  3.4032(39) &  3.4019(14) &  7.8(1) &   3.4046(10) & 15(1)     & 3.41  & 7.8(4)& 3.4018(14)    & 7.8(1) \\
&60   &  0.2315(11) & 0.23112(19) & 33.5(5) &  0.23132(12) & 84(5)    & 0.23  & 31(2)   & 0.23126(19)   & 33.5(5)\\
0.3&30&  21.598(15) & 21.5912(37) & 16.4(1) &  21.5984(21) & 51(1)     & 21.63 & 16.8(4) & 21.5973(39)   & 14.8(2)\\
&50   &   7.114(11) &  7.1169(35) & 9.9(5) &   7.1159(21) & 27(1)   & 7.12  & 11(2)   & 7.1146(35)    & 9.9(1)\\
&60&     3.4954(83) &  3.4514(21) & 15.6(1) &   3.4523(14) & 35(1)     & 3.45  & 15.2(4) & 3.4508(22)    & 14.2(1)\\
\end{tabular}
\end{ruledtabular}
\end{table*}

\begin{table*}
\caption{Same as Table \ref{tablecall} for a
European Put option.}
\begin{ruledtabular}\label{tableput}
\begin{tabular}{ccccccccccc}
        &     & Crude MC& \multicolumn{2}{c}{LSIS($\tilde\mu$)} & \multicolumn{2}{c}{LSIS$(\tilde\mu,\tilde\sigma)$} &\multicolumn{2}{c}{RM}&\multicolumn{2}{c}{GHS}\\
$\sigma$ & $K$ & Price                & Price                 & VR    & Price                 & VR & Price & VR & Price                & VR \\
\hline
0.1&40& 0.004235(98) & 0.0041650(47) & 435(6)  & 0.0041616(41) & 571(9)  & 0.0042& 350(24) & 0.0041650(47)&435(6)  \\
   &50&   0.9636(19) & 0.96419(64)   & 8.8(1)   & 0.96436(38)   &  25(2)   & 0.97  & 9.6(4) & 0.96385(63)  & 9.1(1) \\
   &60&   7.3052(46) & 7.3059(19)    & 5.9(1)  & 7.3056(11)    &  17(1)   & 7.31  & 6.3(4)  & 7.3047(19)   & 5.9(1) \\
0.3&30&  0.13397(83) & 0.13445(13)   & 41(1)    & 0.13448(10)  &  69(2)   & 0.13  & 38(4)  & 0.013440(13) & 40.8(5) \\
&   50&   4.6794(65) & 4.6761(27)    & 5.8(1)  & 4.6767(16)    &  16.5(5) & 4.68  & 6.2(4)  & 4.6761(27)   & 5.8(1) \\
&  60 &  10.5203(97) & 10.5236(44)   & 4.9(1)   & 10.5266(26)   &  13.9(2) & 10.54 & 4.8(4)  & 10.5223(46)  & 4.4(1) \\
\end{tabular}
\end{ruledtabular}
\end{table*}

\section{Gaussian Framework}

The Importance Sampling approaches proposed in the literature
usually apply or are generally formulated within a Gaussian
setting, i.e., in a context where the distribution $P(Z)$ defining
the expectation value (\ref{expectproblem}) is a
$d$-dimensional standard normal distribution. For example, this is the case
for a Monte Carlo simulation of a vector diffusive
process of the form \cite{GlassMCbook}
\begin{equation}\label{sde1}
dX(t) = \mu(X(t),t)\,dt+\sigma(X(t),t)\,dW_t~,
\end{equation}
e.g., for the calculation of the price of an option depending
on the path followed by $X(t)$ within a certain time interval
$[0,T]$. Here the precess $X(t)$ and the drift $\mu(X,t)$ are both
$L$-dimensional real vectors, $W_t$ is a $N$-dimensional standard
Brownian motion, and the volatility, $\sigma(X,t)$, is a $L\times N$
real matrix. 

Continuous time processes of the form (\ref{sde1}) are
typically simulated by sampling $X(t)$ on a discrete grid of
points, $0=t_0 < t_1 < \ldots < t_M =T$, by means, for instance,
of a Euler scheme \footnote{The use of other discretization
schemes does not alter the present discussion.}
\begin{equation}\label{sdeeuler}
X_{i+1} = X_i + \mu(X_i,t)\,\Delta t_i+\sigma(X_i,t)\,\sqrt{\Delta
t_i}\, Z_{i+1}~,
\end{equation}
where $X_i = X(t_i)$, $\Delta t_i = t_{i+1}-t_i$, and $Z_{i+1}$ is
a $N$-dimensional vector of independent standard normal variates.
In this representation, each discretized path for the vector
process $X(t)$ can be  put into a one to one correspondence with a
set of $d = N\times M$ independent standard normal variables $Z$.
Hence, the original problem of evaluating the expectation value of
a functional of the realized path of the process $X(t)$ can be
formulated as in (\ref{expectproblem}). More precisely, $G(Z)$ is
given by the discretized payout functional, and the probability
density is given by a $d$-dimensional standard normal distribution
\begin{equation}\label{gaussmulti}
P(Z) = {(2\pi)^{-d/2}} \,\, e^{ -Z^2/2}~,
\end{equation}
where $Z^2 = Z\cdot Z$.

\begin{table*}
\caption{\label{butterfly} European Butterfly spread option
(\ref{butteq})  on a lognormal asset (\ref{lognormal}) for
different values of the spot price: comparison between the saddle
point approach of Ref.~\cite{GlassImportSampl99} (GHS), and LSIS
with the optimization of both drift and volatility
[LSIS$(\tilde\mu,\tilde\sigma)$] The parameters used are $r = 0.1$, $K_1=45$,
$K_2=50$, $K_3=55$, $T=1.0$, $\sigma=0.3$, and the number of
simulated paths is 1,000,000.}
\begin{ruledtabular}
\begin{tabular}{cccccc}
       & Crude MC&  \multicolumn{2}{c}{GHS}  &\multicolumn{2}{c} {LSIS$(\tilde\mu,\tilde\sigma)$} \\
 $X_0$ & Price                & Price                 & VR    & Price                & VR \\
\hline
  30  & 0.15818(69)  &  0.15732(33)  & 4.4(8) & 0.157659(40)& 298(5) \\
  40  & 0.4871(11)  &   0.48666(98)  & 1.26(1) &  0.48703(11) & 100(5) \\
  50  & 0.6274(13)  &   0.6274(13)   & 1.00(1) &  0.62750(11) & 140(1) \\
  60  & 0.5156(12)   &  0.5157(10)   & 1.44(1) &  0.515636(93)& 166(2) \\
  70  & 0.32961(97) &  0.32875(67)   & 2.10(1) & 0.329281(73)& 177(2) \\
\end{tabular}
\end{ruledtabular}
\end{table*}

\begin{table*}
\caption{\label{glassasian} Comparison between LSIS and the saddle
point approach of Ref.~\cite{GlassImportSampl99} (GSH): price of
Asian options (\ref{asian})  on a lognormal asset
(\ref{lognormal}) for different values of the volatility $\sigma$,
the strike $K$, and of the number of observation dates $M$. The
parameters used are $r = 0.05$, $X_0=50$, $T=1.0$, and the number
of simulated paths is 1,000,000.}
\begin{ruledtabular}
\begin{tabular}{cccccccc}
 &          &     & Crude MC               & \multicolumn{2}{c}{LSIS}           & \multicolumn{2}{c}{GHS}\\
 $M$& $\sigma$ & $K$ & Price                  & Price                   & VR       & Price                  & VR \\
\hline
 16& 0.1     &  45 & 6.0565(29)             & 6.05522(88)             & 10.86(5) & 6.05537(89)            & 10.62(5) \\
  &         &  50 & 1.9198(22)             & 1.91994(80)             & 7.56(5)  & 1.91914(83)            & 7.03(5)  \\
  &         &  55 & 0.20272(74)            & 0.20235(16)             & 21.4(2)  & 0.20237(16)            & 21.4(2)  \\
 16& 0.3     &  45 & 7.1545(77)             & 7.1531(26)              & 8.8(1)   & 7.1529(27)             & 8.13(1)  \\
  &         &  50 & 4.1730(63)             & 4.1714(20)              & 9.9(1)   & 4.1712(21)             & 9.0(1)   \\
  &         &  55 & 2.2135(48)             & 2.2115(13)              & 13.6(7)  & 2.2116(14)             & 11.8(7)  \\
 64& 0.1     &  45 & 5.9967(28)             & 5.99510(87)             & 10.4(5)  & 5.99500(85)            & 10.9(5)  \\
  &         &  50 & 1.8467(21)             & 1.84522(78)             & 7.2(3)   & 1.84525(81)            & 6.7(3)   \\
  &         &  55 & 0.17519(67)            & 0.17448(14)             & 23(2)    & 0.17443(14)            & 23(1)    \\
 64& 0.3     &  45 & 7.0257(75)             & 7.0204(25)              & 9.0(2)   & 7.0204(26)             & 8.3(2) \\
  &         &  50 & 4.0271(61)             & 4.0222(19)             & 10.3(5)  & 4.0220(20)             & 9.3(5) \\
  &         &  55 & 2.0849(46)             & 2.0794(13)              & 12.5(5)  & 2.0794(13)             & 12.5(5) \\
\end{tabular}
\end{ruledtabular}
\end{table*}

As a prototypical example of exotic option pricing problem,
treated as a test case for the most recently proposed Importance
Sampling strategies
\cite{GlassImportSampl99,SUFU00,SUFU02,Arouna03}, in the following
we will consider different arithmetic Asian style options under a
Black-Scholes log-normal model \cite{BaxterRennie}. In this
case, the underlying asset can be simulated on the time
grid relevant for the payout function by means of the exact
recursion
\begin{equation} \label{lognormal}
X_{i+1} = X_i \exp{ \left[(r-\sigma^2/2)\Delta t_i + \sigma
\sqrt{\Delta t_i} Z_{i+1}\right]}~,
\end{equation}
where $r$ is the risk free interest rate, $\sigma$ is the constant
$N\times L$ real volatility matrix, and $\sigma^2$ is a
$L$-dimensional vector of components $\sigma^2_k = \sum_{j=1}^N
\sigma_{kj}$. For an arithmetic Asian style claim, the discounted
payout function is of the form
\begin{equation}\label{asian}
G(Z) = e^{-rT} \Phi(\bar X)
\end{equation}
where $\bar X =(1/M) \sum_{i=1}^{M} X_i$, and $\Phi(\bar X)$ is
some function of $L$ variables. Clearly, European style options
are recovered in the special case $M=1$.

\begin{table*}
\caption{\label{glassasian2} Importance Sampling plus
Stratification. Comparison between LSIS, and the GHS approach of
Ref.~\cite{GlassImportSampl99} for the Asian Option of Table
\ref{glassasian}. }
\begin{ruledtabular}
\begin{tabular}{ccccccc}
 &          &     & \multicolumn{2}{c}{LSIS+ SS}  &\multicolumn{2}{c}{GHS + SS}\\
 $M$& $\sigma$ & $K$  & Price & VR                    & Price                  & VR \\
\hline
16& 0.3 &  45  & 7.15284(25) & 950(20)   & 7.15266(24) & 1030(10) \\
 &     &  50  & 4.17122(18) & 1225(15)  & 4.17118(18) & 1225(30) \\
 &     &  55  & 2.21183(11) & 1900(100) & 2.21183(11) & 1900(50) \\
 64& 0.3     &  45  & 7.02075(23)  & 1060(30) & 7.02076(23) & 1060(30) \\
  &         &  50  & 4.02251(17) & 1290(30)  & 4.02250(17)& 1290(30) \\
  &         &  55  & 2.07967(13) & 1320(100)  & 2.07965(12)& 1470(100) \\

\end{tabular}
\end{ruledtabular}
\end{table*}

\subsection{Gaussian Trial Distributions}
\label{gtd}

The simplest possible strategy for Importance Sampling in a
Gaussian framework is to try to guide the sampled paths towards
the most important regions of the configuration space (i.e.,
where the contribution of the integrand is the largest), by means
of a change of the drift terms of the process (\ref{sde1}) or
(\ref{sdeeuler}). The corresponding  trial probability density
reads
\begin{equation}\label{trialgauss}
\tilde P_{\tilde\mu}(Z) = (2\pi)^{-d/2}  \,\, e^{-(Z-\tilde\mu)^2/2}~,
\end{equation}
 where $\tilde \mu$ is a
$d$-dimensional vector, and the weight function, as also expected
from the Girsamov theorem \cite{Musiela}, is
\begin{equation}
W_{\tilde\mu} (Z) = \exp\left[- \tilde\mu\cdot Z + \tilde\mu^2/2 \right]~.
\end{equation}

A variety of approaches for the determination of the drift vector
$\tilde\mu$ minimizing the variance of the estimator
(\ref{varnew}) has been recently proposed in the literature
\cite{VasDuf98,GlassImportSampl99,
GlassImportSampl99hjm,SUFU00,SUFU02,Arouna03}. These can be
roughly classified into two families depending on the strategy
adopted. The first strategy, common to the so-called adaptive
Monte Carlo methods, aims to determine the optimal drift vector
though a stochastic minimization of the variance. Such
minimization differs in details in the various methods but always
involves an iterative procedure,  to be performed in a preliminary
Monte Carlo simulation.

In particular, Su and Fu \cite{SUFU00,SUFU02}, building upon
previous work by Vazquez-Abad and Dufresne \cite{VasDuf98}, used a
gradient-based stochastic approximation, dubbed infinitesimal
perturbation analysis, in order to estimate the optimal {\em
uniform shift} of the drift for the diffusion (\ref{sdeeuler}),
minimizing the variance of the estimator (\ref{varnew}). In the
notation of this Section, this translates in working with a trial
density of the form (\ref{trialgauss}) where the drift vector
$\tilde\mu$ has components all equal to a single optimization
parameter. The improvement of this method with respect to the one
of Ref.~\cite{VasDuf98}, is that the minimization is carried out
under the original probability measure (as we also do for LSIS),
while in the latter the minimization was formulated under
the trial probability measure. As a result, the stochastic
minimization applies also for non differentiable payout, thus
making the approach more general. The application of this
technique to partial average Asian options in a Black-Scholes
market, and to caplets under the Cox-Ingersoll-Ross model provides
significative variance reductions \cite{SUFU00,SUFU02}.

Along similar ideas, Arouna \cite{Arouna03} has recently proposed
a different stochastic optimization method for the determination
of  the optimal sampling density (\ref{trialgauss}). Here, in
contrast to the previous approach, all the components of the drift
vector are independently optimized. The method relies on  a
truncated version of the Robbins-Monro algorithm that is shown to
converge asymptotically to the optimal drift, and to provide an
effective variance reduction in a variety of cases. However, as
remarked  by the same author, a critical aspect of the practical
implementation of the Robbins-Monro algorithm is that it depends
on the size of the iterative step. Hence, a particular care needs
to be taken in order for the algorithm to be efficient \footnote{B.
Arouna, {\em private communication}.}.

On the other hand, the second strategy, proposed by Glasserman, Heidelberger, and
Shahabuddin (GHS) \cite{GlassImportSampl99}, relies on a saddle point
approximation to minimize the variance of the estimator
(\ref{varnew}), or equivalently of its second moment (in the
original measure)
\begin{equation}\label{secondmom}
m_2(\tilde \mu) = \int_D \hspace{-1mm} dZ \,\, W_{\tilde \mu}(Z)\,G(Z)^2 \,P(Z)~.
\end{equation}
In fact, if the payout function $G(Z)$ is positive definite, by
defining $F(Z) = \log G(Z)$ one can approximate
Eq.~(\ref{secondmom}) with the zero-order saddle point expansion
\begin{eqnarray}
&&(2\pi)^{-d/2} \int_D \hspace{-1mm}dZ \,\exp\left[2F(Z) - \tilde\mu \cdot Z + \tilde\mu^2 /2  - Z^2/2\right] \nonumber \\
&\simeq& C\,\exp \Big[\max_{Z\in D}\left(2F(Z) - \tilde\mu \cdot Z +  \tilde\mu^2 /2  -
Z^2/2 \right)\Big]~, \nonumber
\end{eqnarray}
where $C$ is a constant. As a result, within this approximation,
the problem of determining the optimal change of drift boils down
to finding the vector $\mu$ such that
\begin{equation}
\max_{Z\in D}\left(2F(Z) - \tilde\mu \cdot Z +  \tilde\mu^2 /2  -
Z^2/2 \right)
\end{equation}
is minimum.  It is easy to show that this is obtained by choosing
$\tilde \mu^\star = Z^\star$ where $Z^\star$ is the point that
solves the optimization problem
\begin{equation} \label{optimization}
\max \left(F(Z) - Z^2/2\right)~,
\end{equation}
or equivalently, for which the payout times the original
distribution, $ G(Z) P(Z)$, is maximum, i.e., $Z^\star$
corresponds to the maximum of the optimal sampling density,
Eq.~(\ref{zerovar}).  The simplest interpretation of the saddle
point approach is therefore that it approximates the zero variance
distribution by means of a normal density with the same mode and
variance.

The saddle point approach can be expected to be particularly
effective in reducing the variance of the Monte Carlo estimator
whenever the log payout function $F(Z)$ is close to be linear in
the portion of the configuration space where  most of the
probability mass of $P(Z)$ lays. Moreover, GHS have also shown
that stratifying \cite{GlassMCbook} the $d$-dimensional random
vector $Z$ along the optimal $\mu$, provides a spectacular
variance reduction in certain cases
\cite{GlassImportSampl99,GlassImportSampl99hjm}. However, whenever
the optimal sampling probability (\ref{zerovar}) cannot be
accurately represented by a single Gaussian with the same mode and
variance, the saddle point approximation is less beneficial. In
particular, as it will be also illustrated in
Sec.~\ref{numexample}, this approach turns out to be less
effective whenever the structure of the payout function $G(Z)$ is
such that the optimal sampling distribution (\ref{zerovar}) has a
width  which is very different from the one of the original
distribution, or is multi-modal.

In the following Section we describe an alternative least squares
strategy that is straightforward to implement and flexible enough
to be applied in a generic Monte Carlo setting. Indeed, the Least
Squares Importance Sampling (LSIS) is not limited to the
determination of the optimal change of drift in a Gaussian
model. Instead, it can be applied to any Monte Carlo simulation
provided that a reasonable guess of the optimal sampling density
is available. For this reason, in the next Section we will
momentarily leave the Gaussian framework, and we will describe the
rationale of LSIS in a more general setting.

\section{Least Squares Importance Sampling}
\label{leastsquares}

A practical approach to the search of an effective Importance
Sampling distribution can be formulated in terms of a non-linear
optimization problem. To this purpose, let us consider a family of
trial probability densities, $\tilde P_{\theta}(Z)$, depending on a
set of $N_\theta$ real parameters $\theta = (\theta_1, \theta_2,
\ldots, \theta_{N_\theta})$~. The variance of the estimator
corresponding to $\tilde P_{\theta}(Z)$, Eq.~(\ref{varnew}), can
be written in terms of the {\em original} probability distribution
$P(Z)$ as
\begin{equation}\label{varest}
\tilde \Sigma_\theta^2 = E_P \left[W_\theta(Z) G^2(Z) \right] -
E_P \left[G(Z) \right]^2~,
\end{equation}
with $W_\theta(Z) = P(Z)/\tilde P_\theta(Z)$. Hence, the optimal
Importance Sampling distribution within the family $\tilde P_\theta(Z)$
is the one for which the
latter quantity, or equivalently the second moment
(\ref{secondmom}) or
\begin{equation}\label{second}
E_P \left[W_\theta(Z) G^2(Z) \right]~,
\end{equation}
is minimum. The crucial observation is that the Monte Carlo
estimator of this quantity,
\begin{equation}\label{minimizer}
m_2(\theta) \simeq \frac{1}{N_p^\prime} \sum_{i=1}^{N_p^\prime} \left( W_\theta(Z_i)^{1/2}
G(Z_i)\right)^2 \,\,\,\,\, Z_i\sim P(Z)~,
\end{equation}
can be interpreted as a non-linear least squares fit of a set of
$N_p^\prime$ data points $(x_i,y_i)$ with a function $y = f_\theta(x)$
parameterized by $\theta$, with the correspondence $y_i \to 0$,
$x_i \to Z_i$, and $f_\theta(x) \to W_\theta(Z)^{1/2} G(Z)$.
The latter is a standard problem of statistical analysis that can
be tackled with a variety of robust and easily accessible
numerical algorithms, as the so-called Levenberg-Marquardt method
\cite{numrec}.

Alternatively, to improve the numerical stability of the
least-squares procedure, it is convenient in some situations
to minimize, instead of (\ref{second}),  the pseudo-variance
\begin{eqnarray}\label{pseudovariance}
S_2(\theta) &=& E_P\left[ \left( W_\theta(Z)^{1/2}
G(Z) - V_T \right)^2\right] \nonumber \\
&&\simeq\frac{1}{N_p^\prime}\sum_{i=1}^{N_p^\prime} \left( W_\theta(Z_i)^{1/2}
G(Z_i) - V_T \right)^2 \,
\end{eqnarray}
where the constant $V_T$ is a guess of the option value. Indeed, the minimization
of (\ref{pseudovariance}) is equivalent to the one of the real variance of the estimator
(\ref{varest}) as
\begin{equation}
S_2(\theta) = \tilde \Sigma_\theta^2 + \left(E_P\left[G(Z)\right] - V_T\right)^2~.
\end{equation}

The algorithm for the determination of the optimal
sampling distribution within a certain trial family can be therefore
summarized as it follows:
\begin{itemize}
\item Generate a suitable number $N_p^\prime$ of replications of the
state variables $Z$ according to the {\em original} probability
distribution $P(Z)$;

\item Choose a trial probability distribution $\tilde P_\theta(Z)$, and
an initial value of the vector of parameters $\theta $;

\item Set $ x_i\to Z_i$, $f_\theta(x) \to W_\theta(Z)^{1/2} G(Z)$
and $y_i \to 0$ ({\em resp.} $y_i \to V_T$) and call a least squares fitter, say ${\rm
LSQ}\,\left[x,y, f_\theta(X),\theta\right]$~, providing the
optimal $\theta= \theta^\star$ by minimizing the second moment of
the estimator $m_2(\theta)$, Eq.~(\ref{minimizer}) [{\em resp.}
$S_2(\theta)$, Eq.~(\ref{pseudovariance})].

\end{itemize}

Once the optimal parameters $\theta^\star$ have been determined through the least squares algorithm, one can
perform an ordinary Monte Carlo simulation  by sampling the probability distribution
$\tilde P_{\theta\star}(Z)$, and calculating expectation values according to Eq.~(\ref{isamp}).

As it will be illustrated with the numerical examples of
Sec.~\ref{numexample}, this procedure does not add a significant
overhead to the simulation, because just a relatively small number
of replications $N_p^\prime\ll N_p$ is usually required to
determine the optimal parameter $\tilde \theta^\star$. This is due
to the fact that the configurations over which the optimization is
performed are fixed. As a result of this form of {\em correlated
sampling} \cite{umrigar98}, the difference in the $m_2(\theta)$'s
for two sets of values of the parameters being optimized is much
more accurately determined than the values of the $m_2(\theta)$'s
themselves. This rather surprising feature is rooted in the fact
that the minimization of Eq.~(\ref{minimizer}) as a means to
optimize the trail density, $\tilde P_\theta(Z)$, can  be
justified in terms of a genuine maximum likelihood criteria
\cite{Bressanini02}, and it is therefore independent on how
accurately $m_2(\theta)$ approximates the quantity (\ref{second}).

Clearly, the fact that only a limited number
of Monte Carlo samples is required for
the optimization of the trial density limits the overhead
introduced by the LSIS algorithm. This makes LSIS a practical
strategy for variance reduction.

In the following we will demonstrate the effectiveness of this
simple strategy by applying it to a variety of test cases, and we
will compare the present approach with the ones most recently
proposed in the literature
\cite{GlassImportSampl99,SUFU00,SUFU02,Arouna03}.

\begin{figure}
\vspace{1mm}
\includegraphics[width=0.43\textwidth]{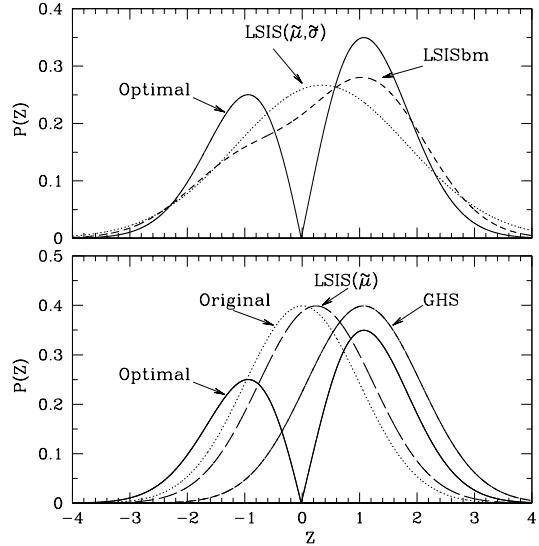}
\vspace{-3mm} \caption{\label{figstradd} Sampling probability
density functions for a straddle option (\ref{payoutstradd}) with
$r = 0.05$,  $X_0 = K = 50$ , $\sigma = 0.3$, $T = 1$, as obtained
with LSIS [optimizing just the drift, LSIS$(\tilde\mu)$,  both the
drift and the volatility , LSIS$(\tilde\mu,\tilde\sigma)$, and
using the bi-modal trial density (\ref{bimodal}), LSISbm], and the
saddle point approximation (GHS). The original (\ref{gaussmulti})
and the optimal (\ref{zerovar}) sampling densities also shown for
comparison.}
\end{figure}

\section{Numerical Examples}\label{numexample}

\subsection{European Options}
\label{europeanexamples}

We will start by considering European style options. These are the
simplest possible examples of financial relevance that allow one
to illustrate the LSIS strategy. In particular, we will
consider first standard Call and Put options, written on a single
underlying asset, $X(t)$, following the log normal process
(\ref{lognormal}). Here $N = M = L = 1$, and $\Delta t_0 = T$. The
payout function reads as in Eq.~(\ref{asian}) with $\bar X \to X_1
\equiv X(T)$, $\Phi(X) = (X-K)^{+}$ and $\Phi(X) = (K-X)^{+}$ for
the Call and the Put, respectively, and $K$ is the strike price.

In these cases, the probability distribution $P(Z)$ is a
univariate Gaussian distribution, and the option value
(\ref{expectproblem}) involves the integration of a function of a
single variable. As discussed in Sec.~\ref{is},  Importance
Sampling techniques seek a sampling probability density $\tilde
P_\theta(Z)$ as close as possible to the optimal sampling
distribution, Eq.~(\ref{zerovar}) (see Figure \ref{figcall}). The
simplest choice for $\tilde P_\theta(Z)$, in this setting, is a
Gaussian distribution of the form (\ref{trialgauss}) (with $d=1$),
so that the only parameter $\theta$ to optimize is the drift
$\tilde\mu$.

The LSIS algorithm described in Sec. \ref{leastsquares} is
particularly straightforward to implement, and involves a
pre-simulation stage in which a set of $N_p^\prime$ standard
normal random numbers are generated and stored, and a least
squares optimization routine is invoked. As anticipated, we found
that the least squares fitter was able to determine successfully
the optimal $\tilde \mu$ with as little as $N_p^\prime \simeq 50$
Monte Carlo replications, thus making the approach useful in
practice.

In Tables \ref{tablecall} and \ref{tableput} we compare the
results obtained with LSIS
with the ones obtained by means of the Robbins Monro (RM) adaptive
Monte Carlo (as quoted in Ref.~\cite{Arouna03}), and the saddle
point approach of GHS \cite{GlassImportSampl99}.
 Here,  as an indicator of the efficiency gains
introduced by the different strategies of Importance Sampling, we have defined the
variance ratio as
\begin{equation}\label{varred}
{\rm VR} = \left(\frac{\sigma({\rm Crude\,\,MC})}{\sigma({\rm IS})}\right)^2
\end{equation}
where the numerator and denominator are the statistical errors (for the same number of Monte Carlo paths)
of the Crude and the Importance Sampling estimators, respectively.
In particular, for this and the following examples, we have estimated the uncertainties
on the statistical errors, and on the VR's by performing a standard
error analysis of the outcome of several Monte Carlo simulations
for different random number seeds.

We found that the different methods produce a significative and
comparable variance reduction. In particular, as it is intuitive,
the change of drift is more effective for low volatility,
and deep in and out of the money options. In this case,  LSIS and
GHS optimized trial distributions $\tilde P_{\bar \mu}(Z)$ are
very similar as shown Fig.~\ref{figcall}. This could be expected
as, in this case, the optimal Importance Sampling  distribution
(\ref{zerovar}) can be effectively approximated by a Gaussian with
the same mode and variance, so that the GHS approach produces
accurate results.

\begin{table*}
\caption{\label{tableasianRM} Comparison between LSIS, the
adaptive Robbins-Monro (RM) method  (as quoted in
Ref.~\cite{Arouna03}), for the price of an Asian Call option
(\ref{asian}) on a lognormal asset (\ref{lognormal}) for different
values of the volatility $\sigma$, and of the strike price $K$. The
parameters used are $r = 0.1$, $X_0=50$, $T=0.5$, and the number
of simulated paths is 1,000,000 for Crude MC and LSIS, 800,000 for
RM. The uncertainties on the RM variance reductions are rough estimates based on
the number of simulated paths.}
\begin{ruledtabular}
\begin{tabular}{cccccccc}
  &            &     &  Crude MC& \multicolumn{2}{c}{LSIS}   & \multicolumn{2}{c}{RM}    \\
$M$ &  $\sigma$  & $K$ & Price                 & Price & VR   &
Price   & VR  \\
\hline
20&  0.1 &  40 & 10.7820(21)  & 10.78243(22) & 91(2)   & 10.73   & 58(2)  \\
  &      &  50 &  1.6045(16)  & 1.60398(61)  & 6.9(1)  & 1.53    & 7.0(1)   \\
  &      &  55 &  0.04919(29) & 0.049204(54) & 28.8(5)   & 0.037   & 6.0(5)   \\
  &  0.3 &  40 &  10.8284(62) & 10.8277(17)  & 13(1)   & 10.77   & 14(1)  \\
  &      &  50 &  3.1373(44)  & 3.1352(15)   & 8.6(1)  & 2.99    & 9.0(1)   \\
  &      &  60 &  0.3790(16)  & 0.37861(35)  & 20.9(5) & 0.320    & 28.0(5)  \\
\end{tabular}
\end{ruledtabular}
\end{table*}

However, the LSIS method is not limited to Importance Sampling
strategies based on a pure change of drift, and one can easily
introduce additional optimization parameters  in the trial density.
For instance, in this example it makes sense to introduce the
sampling volatility,  $\tilde \sigma$,
\begin{equation}\label{trialvol}
\tilde P_{\tilde\mu,\tilde\sigma}(Z) = (2 \pi \tilde \sigma^2)^{-1/2}
e^{-(Z-\tilde\mu)^2/2\tilde \sigma^2}~.
\end{equation}
As illustrated in Fig.~\ref{figcall}, by adjusting both $\tilde\mu$ and
$\tilde \sigma$, one obtains a trial density closer to
the optimal one. This corresponds to an additional variance
reduction up to over one order of magnitude, as shown in Tables
\ref{tablecall} and \ref{tableput}.

Another simple example in which the ability to adjust the width of
the sampling distribution proves effective is the Butterfly spread
option,
\begin{equation}\label{butteq}
\Phi(X) = (X-K_1)^+ - 2 (X-K_2)^- + (X-K_3)^+~,
\end{equation}
with $K_1<K_2<K_3$. In this case, a pure change of drift can only
force the expected end point of the sampled paths $X(T)$ to fall
in the money, $K_1 < X(T) < K_3$. However, for values of the
volatility or the maturity large enough, a significative fraction
of the sampled paths will still results in a zero payout. Hence, a
pure change of drift is not very effective in reducing the
variance. This is illustrated in Table \ref{butterfly}. Instead,
by quenching the sampled volatility one can increase the number of
paths falling in the money thus reducing the variance. For
instance, in the example of Table \ref{butterfly}, for $S_0 = 50$
the saddle point change of drift is very small, and does not alter
the variance. Instead, by allowing the sampling volatility to
vary, we get $\tilde \mu^\star \simeq
 - 0.02$ and $\tilde\sigma^\star \simeq  0.14$, reducing
the variance by more than two orders of magnitude.

\begin{figure}[h]
\includegraphics[width=0.34\textwidth, angle = 270]{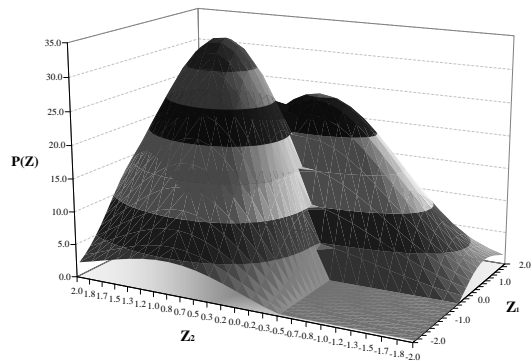}
\vspace{-3mm} \caption{\label{figpuppa} Optimal sampling distribution, Eq.~(\ref{zerovar}),
for the European basket Call option (\ref{basket}) for $r=0.05$, $K=100$, $X_1^0 = 100$, $X_2^0 =
105$, $\sigma_1=\sigma_2 = 0.3$, $T=1$.  }
\end{figure}

\begin{table*}
\caption{\label{tableasianSF} Comparison between LSIS, the
adaptive approach of Su and Fu (SF) (as quoted in
Ref.~\cite{SUFU00}), for the price of a partial average Asian Call option
(\ref{asian}) and (\ref{pavg}) on a lognormal asset
(\ref{lognormal}) for different values of the volatility $\sigma$,
and of the strike price $K$. The parameters used are $r = 0.05$,
$X_0=100$, $T=1.0$, $M=365$, $M_0=305$, and the number of
simulated paths is 1,000,000 for Crude MC, and LSIS, and 50,000
for SF. The uncertainties on the SF variance reductions are rough estimates based
on the number of simulated paths.}
\begin{ruledtabular}
\begin{tabular}{ccccccc}
            &     &  Crude MC             & \multicolumn{2}{c}{LSIS}  & \multicolumn{2}{c}{SF}    \\
  $\sigma$  & $K$ & Price                 & Price                & VR & Price            & VR  \\  \hline
  0.2       & 100 & 9.792(14)     &  9.7816(46)   & 9.3(1)   & 9.747 & 6.7(3) \\
            & 110 & 5.413(11)     &  5.4256(33)   & 11.1(1)  & 5.397 & 8.2(3) \\
            & 120 & 2.7758(77)    &  2.7669(20)   & 14.8(2)  & 2.730 & 11.0(6) \\
            & 130 & 1.3038(52)    & 1.3096(12)    & 18.8(1)  & 1.284  & 17.0(4)  \\
            & 140 & 0.5861(35)    & 0.58298(61)   & 33(1)    & 0.575  & 25(3)  \\
            & 150 & 0.2421(22)    & 0.24592(30)   & 54(4)    & 0.241  & 44(12)  \\
            & 160 & 0.1003(14)    &   0.09973(14) & 100(8)   & 0.0980 & 85(20)  \\
            & 170 & 0.03761(84)   &  0.039000(64) & 172(20)  & 0.0380 & 173(80) \\
  0.3       & 100 & 13.323(21)    & 13.3527(65)   & 10.4(5)  & 13.295& 7(1) \\
            & 110 & 9.182(18)     & 9.1564(54)    & 11.1(5)  & 9.103 & 8(1) \\
            & 120 & 6.1041(15)    & 6.1258(39)    & 14.8(5)  & 6.059 & 10(1) \\
            & 130 & 4.020(12)     &  4.0034(31)   & 15.0(5)  & 3.985 & 12(3) \\
            & 140 & 2.5624(98)    &  2.5769(23)   & 18.1(7)  & 2.556 & 15(2) \\
            & 150 & 1.6410(80)    & 1.6325(13)    & 38(1)    & 1.603 & 22(2) \\
            & 160 & 1.0143(62)    &  1.02355(90)  & 47(2)    & 1.006 & 30(5) \\
            & 170 & 0.6408(50)    &  0.63681(61)  & 67(3)    & 0.623 & 36(6) \\
\end{tabular}
\end{ruledtabular}
\end{table*}

\subsection{Asian Call option}

As a second example we consider here the single underlying ($L=1$)
arithmetic Asian option Eq.~(\ref{asian}), with a Call 
payout, $\Phi(Z) = e^{-rT}\, (\bar X - K)^+$, and $M$ observation
dates.

In this case, the calculation of the expectation value
(\ref{expectproblem}) involves a multi-dimensional integral over
the probability distribution (\ref{gaussmulti}). However, the
implementation of LSIS is not more difficult than  for the
European payout. Indeed, the fact that the sample points $Z_i$ are
in the present case $M$-dimensional vectors is practically
irrelevant for the LSIS procedure. All the least square fitter
needs is a method  that, given a {\em configuration} $Z_i$ and the
vector $\theta$, provides the value of the target function
$f_\theta(Z_i) = W_\theta(Z_i)^{1/2}G(Z_i)$.

In order to compare LSIS  with the other approaches proposed
in the literature we have considered a Gaussian trial density of
the form Eq.~(\ref{trialgauss}). Here such density depends on as
many parameters, $\tilde \mu = (\tilde \mu_0,\ldots,\tilde \mu_{M-1})$, as are the sampling
dates in the Euler discretization of the process (\ref{sdeeuler}).
A typical outcome of the optimized drift vector produced by the LSIS
procedure is shown in Fig.~\ref{figmu}.

\subsubsection{Comparison with the Saddle Point Approximation}

As shown in Table \ref{glassasian}, LSIS provides a very effective
variance reduction for the Asian Call options considered, similar to the one obtained with the approach
of GHS. This could be
expected on general grounds as in the case of the European Call and Put options.
Indeed, it is not difficult to verify that the single asset Asian
Call option generates a single-mode optimal sampling density $P_{\rm opt}(Z) \propto
G(Z)P(Z)$, with covariance close to one. Hence, as discussed in
the previous Section, the optimal sampling density can
be accurately represented by a Gaussian distribution, and the
saddle point method provides an accurate approximation of
the optimal drift vector. Indeed, as shown Fig.~\ref{figmu}, the
optimal drift vectors obtained with the LSIS and the GHS
approaches are quite similar, with an overlap of $\sim 0.99$ when
both normalized to one.

\begin{table*}
\caption{\label{straddletable} Importance Sampling results for the
European style straddle (\ref{payoutstradd}) obtained by means of
the saddle point approximation (GHS) and LSIS, with the
optimization of the drift [LSIS($\tilde\mu$)], and the drift and
the volatility [LSIS$(\tilde{\mu},\tilde\sigma)$]. The parameters
used are $r= 0.05$, $X_0=K=50$, $\sigma=0.3$, $T=1$, and
correspond to the probability distributions sketched in
Fig.~\ref{figstradd}. The number of simulated paths is 1,000,000.}
\begin{ruledtabular} .
\begin{tabular}{ccccccc}
    Crude MC            & \multicolumn{2}{c}{GHS}  &\multicolumn{2}{c}{LSIS($\tilde\mu$)}  & \multicolumn{2}{c}{LSIS$(\tilde\mu,\tilde\sigma)$}  \\
    Price               & Price         & VR       & Price & VR                    & Price & VR   \\
\hline
    11.803(10)          & 11.800(33)   &  0.10(1)  & 11.8038(88) & 1.30(1)         &  11.8047(55) &  3.00(2)   \\
\end{tabular}
\end{ruledtabular}
\end{table*}

As anticipated, we found that a few hundreds of Monte Carlo
configurations and 10-20 iterations of the least squares fitter,
were typically enough to determine the optimal drift vector in
(\ref{trialgauss})  for $M \simeq 50$, thus making the overhead of
the pre-simulation stage rather limited. However, as the number of
time steps -- or more in general the number of components of the
drift vector -- increases, the complexity of the optimization problem
increases as well. Nevertheless, as suggested in
Ref.~\cite{GlassImportSampl99hjm}, one can significantly reduce the
computation time associated with the optimization stage by
approximating the drift vector with a continuous function
parameterized by a small number of parameters. These are
in turn tuned by the least square algorithm in order to determine 
an approximate optimal
drift vector. We have found that a particularly
effective realization of this approach is to approximate the
drift vector by a piecewise linear function, parameterized by its
values where it changes slope (the so-called {\em knot points}). For
instance, in the present examples, optimizing over 4 to 8 equally
spaced knot points provides variance reductions practically
indistinguishable from the ones obtained by a full optimization.
In addition, the drift vector generally changes continuously with
the market parameters. As a result, an additional computational
saving can be obtained by starting the pre-simulation from a
drift vector optimized for a case with a similar set of
parameters.

As anticipated in Sec. \ref{gtd}, Importance Sampling can be also
naturally combined with Stratified Sampling by choosing the
optimal drift vector as the stratification direction (for further
details see Ref.\cite{GlassImportSampl99}). This, as shown in
Table \ref{glassasian2}, gives rise to a further reduction of the
variance, which can be of several order of magnitudes, depending
on the option parameters, thus resulting in quite remarkable
savings in computational time.

\subsubsection{Comparison with Adaptive Monte Carlo approaches}

It is also interesting to compare the LSIS results with those
obtained by means of the recently proposed  adaptive Monte Carlo
methods briefly discussed in Sec.~\ref{gtd}.  As illustrated in
Table \ref{tableasianRM}, we found that our approach achieves
variance reductions similar to the ones obtained with the
Robbins-Monro algorithm of Ref.\cite{Arouna03}. However, for the
example considered, LSIS appears to perform better for small
values of the volatility and deep in and out of the money options.
This is particularly remarkable when considering that the
pre-simulation stage required by the Robbins-Monro algorithm
involves sampling a number of Monte Carlo paths much larger than
that required by the LSIS approach. For instance, in the present
example, the number of iterations quoted in Ref.~\cite{Arouna03}
is 400,000 while the LSIS results were obtained by sampling just
$N_p^\prime \simeq 400$ configurations. In addition, LSIS does not
have any exogenous parameter to be fine tuned in order to achieve
an efficient optimization, thus making the approach easier to be
automated.

Finally, in order to compare with the adaptive Monte Carlo
approach of Su and Fu \cite{SUFU00,SUFU02}, we have slightly
modified the Asian option example and considered a payout
depending on a partial average
\begin{equation}\label{pavg}
\bar X = \frac{1}{M-M_0} \sum_{i=M_0}^M X_i~,
\end{equation}
where $M_0$ is the time step index corresponding to the first
observation.
The results we have obtained in this case are reported in
Table \ref{tableasianSF}, and show that LSIS generally outperforms
the approach of Ref.~\cite{SUFU00,SUFU02}. This is expected as LSIS
involves a more general class of sampling distribution
(\ref{trialgauss}) as the change of drift is not constrained to be
uniform along the simulation path.

\subsection{Multi-modal Examples}
\label{multimodalsec}

All the examples considered in the previous Sections were
characterized by a single mode optimal sampling distribution
(\ref{zerovar}). In these cases, an Importance Sampling strategy
based on the Gaussian trial distributions (\ref{trialgauss}) or
(\ref{trialvol}) provides good results. In this Section we will
consider instead option pricing problems that are characterized by
multi-modal optimal sampling densities. These are very common in
practice, and typically arise if the payout function $G(Z)$ is not
monotonic, or for claims written on multiple assets. We will show
that in these cases a simple Importance Sampling strategy based on
a pure change of drift of a Gaussian density proves ineffective.
In contrast, LSIS allows one to work with trial densities tailored
to the problem at hand, including multi-modal distributions.
As in Section \ref{europeanexamples}, we begin by
illustrating the effectiveness of LSIS with a simple
European style payout.

\subsubsection{Straddle}

A simple example of a pricing problem characterized by a
multi-modal sampling density is the European straddle:
\begin{equation}\label{payoutstradd}
\Phi(X) = (X-K)^{+} + (K-X)^{+}~.
\end{equation}
As illustrated in Fig.~\ref{figstradd}, the corresponding optimal
sampling distribution,  $P_{\rm opt}(Z) \propto 
G(Z)\,\exp{(-Z^2/2)}$, has two well separated maxima. As a result,
saddle point approaches based on a Gaussian distribution, e.g.,
centered on the higher of the two modes, provide a poor approximation
of the optimal sampling density. Indeed, as shown in Table \ref{straddletable}, the saddle
point method {\em increases} the variance with respect to the
crude Monte Carlo estimator. On the contrary,  LSIS based on the
simple trial density (\ref{trialgauss}) provides a small shift of
the drift towards the higher mode of the optimal distribution
[Fig.~\ref{figstradd}]. This can be easily understood. In fact, the
best way to approximate a bi-modal distribution by means of a 
Gaussian density is to center
the latter around the center of mass of the former, in this case
$\tilde\mu \simeq 0.24$. Yet, since the two maxima of the optimal
distribution are well separated, its overlap with the trial density
is still rather poor. As a result, LSIS based on a trial 
density of the form (\ref{trialgauss}) produces only a small variance
reduction. Better results can be obtained by adjusting the width
of the trial density. Indeed, as also illustrated in
Fig.~\ref{figstradd}, by increasing the standard deviation of the
sampling density one achieves a better approximation of the
optimal distribution. This leads to a sizable reduction of the
variance of a factor of 3. 

An even better {\em ansatz} for
the optimal density is clearly represented by a bi-modal trial
density of the form
\begin{equation}\label{bimodal}
\tilde P(Z) = (2\pi)^{-d/2}\Big[w_a \,e^{-(Z-\mu_a)^2}+w_b\,
e^{-(Z-\mu_b)^2}\Big]~,
\end{equation}
where $w_a+w_b=1$ (and $d=1$ in this specific case) that can be
optimized over $\mu_a$, $\mu_b$, and $w_a$. The simulation of a
density of this form is straightforward as it simply implies
choosing on each Monte Carlo step one of the two Gaussian
components in (\ref{bimodal}), and sample a configuration $Z_i$
according to it. This can be done by extracting an auxiliary
uniform random number $\xi \in [0,1]$, and extracting $Z_i$
according to the first Gaussian component if $\xi<w_a$, and
according to the second otherwise. The optimized distribution
obtained using this bi-modal trial density is sketched in
Fig.~\ref{figstradd}, and corresponds to a further sizable
reduction of the variance (Table \ref{straddletable}). 
Remarkably, the same form of
sampling density (\ref{bimodal}) can be used also for Asian style
straddles (here $d=M$, $\mu_a$ and $\mu_b$ are $M$-dimensional
vectors), and produces a very effective variance reduction, as illustrated
in Table \ref{straddletable2}.

\begin{table}[h]
\caption{\label{straddletable2} Variance Reduction obtained with
LSIS for a European ($M=1$), and Asian style ($M=16$ and $64$)
straddle (\ref{payoutstradd}) using the bi-modal sampling
distribution (\ref{bimodal}). The parameters used are $r= 0.05$,
$X_0=K=50$, $\sigma=0.3$, $T=1$, and correspond for $M=1$ to the
probability distributions sketched in Fig.~\ref{figstradd}. The
number of simulated paths is 1,000,000.}
\begin{ruledtabular}
\begin{tabular}{cccccc}
  &  Crude MC& \multicolumn{2}{c}{LSIS}  \\
$M$&  Price   & Price & VR            \\
\hline
 1  & 11.803(10)   &  11.8009(44) & 5.17(5)     \\
 16 & 7.0604(58)   &   7.0559(26) & 4.98(5)     \\
 64 & 6.8200(55)   &   6.8163(26) & 4.47(5)     \\
\end{tabular}
\end{ruledtabular}
\end{table}

\subsubsection{Basket Call Options}

Basket options are another very common class of contingent claims
that can give rise to multi-modal optimal distributions. This is
illustrated in Fig.~\ref{figpuppa} for a very simple European
style Call option on the maximum of $L=2$ underlying assets
following the process (\ref{lognormal})
\begin{equation}\label{basket}
\Phi(X) = (\max(X_1,X_2)-K)^+~.
\end{equation}
Also in this case, LSIS based on a bi-modal trial density of the
form (\ref{bimodal}) provides a significative variance reduction
that persists when introducing Asian features in the payout
(see Table \ref{tablemultieuro}). We  obtained similar results
for other multi assets options by generalizing the form of the
trial density (\ref{bimodal}).

\begin{table}
\caption{\label{tablemultieuro} Importance Sampling results for
the Asian style basket option (\ref{basket}) obtained by means of
LSIS, using the bi-modal sampling distribution (\ref{bimodal}). The
parameters used are $r= 0.05$, $T=1$, $X_0^{(1)} = 100$,
$X_0^{(2)} = 105$, $\sigma_1 =0.3$, $\sigma_2 =0.3$, and
correspond (for $K=100$ and $M=1$) to the probability
distributions sketched in Fig.~\ref{figpuppa}. The number of
simulated paths is 500,000.}
\begin{ruledtabular}
\begin{tabular}{ccccccc}
&  &  Crude MC& \multicolumn{2}{c}{LSIS}  \\
$M$ & $K$&  Price   & Price & VR            \\
\hline
1  & 90  &  34.555(41)   &  34.587(14) &  8.58(7)    \\
   & 100 &  26.621(39)   &  26.589(15) &  6.76(3)    \\
   & 110 &  19.749(36)   &  19.776(13) &  7.7(1)    \\
16 & 90  &  24.876(24)   &  24.8700(87)&  7.6(1)     \\
   & 100 &  16.428(22)   &  16.4081(81) &  7.4(1)    \\
   & 110 &  9.711(19)    &  9.6918(68) &  7.8(1)    \\
32 & 90  &  24.537(23)   &  24.5539(84) &  7.50(5)    \\
   & 100 &  16.082(22)   &  16.0734(82) &  7.2(1)    \\
   & 110 &  9.381(18)    &   9.3752(68) &  7.0(1)   \\
\end{tabular}
\end{ruledtabular}
\end{table}

\section{Conclusions}

In this paper we have described a simple Importance Sampling
technique based on a least squares minimization of straightforward
implementation. The resulting strategy, dubbed Least Square
Importance Sampling (LSIS), provides an effective variance reduction
technique that lends itself to a variety of applications.

We have presented several numerical examples in a diffusive
setting, and we have shown that LSIS, when restricted to the
optimization of the drift of a diffusion process, provides
variance reductions similar to those obtained with existing
approaches \cite{GlassImportSampl99,SUFU00,SUFU02,Arouna03}.
However,  LSIS is not limited to the determination of the optimal
mean of  a Gaussian sampling distribution. As a result, it
outperforms other approaches when the ability to adjust
the width of such distribution, or to sample non-Gaussian and
multi-modal densities, is important to achieve variance reductions.

The LSIS strategy can be applied to any Monte Carlo setting
provided that a reasonable ansatz for the optimal sampling
distribution is available. This makes LSIS a flexible Importance
Sampling approach, that can be used across a variety of financial
applications, ranging from Value at Risk (VaR) estimation, to
portfolio credit risk management. This is currently the object of
further investigations.

\acknowledgments It is a pleasure to acknowledge Gabriele
Cipriani, David Shorthouse, and Mark Stedman for stimulating
discussions, and Paul Glasserman for an enlightening lecture that
inspired this work. The opinion and views expressed in this paper
are uniquely those of the author, and do not necessarily represent
those of Credit Suisse Group.

\bibliography{biblio}

\end{document}